# TRANSPORT OF LCLS-II 1.3 GHZ CRYOMODULE TO SLAC

M.W. McGee[†], T. Arkan, T. Peterson, Z. Tang, Fermi National Accelerator Labortory*, Batavia, IL 60510
S. Boo, M. Carrasco, SLAC National Accelerator Labortory, Menlo Park, CA 94025
E. Daly, N. Huque, Thomas Jefferson National Accelerator Facility, Newport News, VA 23606

*Abstract*

In a partnership with SLAC National Accelerator Laboratory (SLAC) and Jefferson Lab, Fermilab will assemble and test 17 of the 35 total 1.3 GHz cryomodules for the Linac Coherent Light Source II (LCLS-II) Project. These include a prototype built and delivered by each Lab. Another two 3.9 GHz cryomodules will be built, tested and transported by Fermilab to SLAC. Each assembly will be transported over-the-road from Fermilab or Jefferson Lab using specific routes to SLAC. The transport system consists of a base frame, isolation fixture and upper protective truss. The strongback cryomodule lifting fixture is described along with other supporting equipment used for both over-the-road transport and local (on-site) transport at Fermilab. Initially, analysis of fragile components and stability studies will be performed in order to assess the risk associated with over-the-road transport of a fully assembled cryomodule.

## INTRODUCTION

The cryomodule transport design acceleration criteria were initially established by considering the 805 km over-the-road Spallation Neutron Source (SNS) transport from Jefferson Lab in Newport News, Virginia to Oak Ridge, Tennessee [1]. A transport analysis completed by Babcock Noell regarding TTF style cryomodules found that the acceleration limits for the input coupler (IC), perpendicular to the antenna, must be less than 1.5 g [2]. Initial transport studies were conducted for XFEL 1.3 GHz Cryomodules to establish the over-the-road transport from CEA-SACLAY to DESY [3]. Most recently, an acceleration limit criteria for XFEL cryomodule transport to DESY was established as 1.5 g (vertical and longitudinal) and 1 g (transversely) [4]. Successful transport of 80 of 100 cryomodules to date have been completed by following this acceleration limit criteria [5].

## CRYOMODULE DESIGN

The overall structural design of the LCLS-II cryomodule is similar to that of the TESLA-style module, shown in Figure 1. It consists of eight dressed 9-cell niobium superconducting radio frequency (RF) cavities with a quadrupole at the downstream end. This coldmass hangs from three column support posts constructed from G-10 fiberglass composite, which are attached to the top of the vacuum vessel. The helium gas return pipe (HeGRP), supported by the three columns acts as the coldmass spine, supporting the cavity string and ancillaries. Support brackets with adjusting blocks using needle bearings on each side provide a connection between each cavity and the HeGRP. Two aluminum heat shields hang from the HeGRP column supports [4].

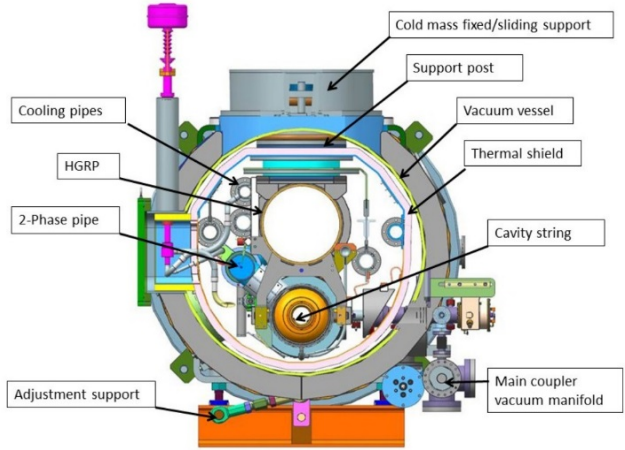

Figure 1: Cross-section of the cryomodule showing its major sub-assemblies [6].

## TRANSPORT ASSEMBLY DESIGN

The transport system consists of a base frame, isolation fixture and upper (protective) truss as shown in Figure 2. The base frame is a welded structural steel truss, capable of lifting the entire assembly through (4) lifting lugs and this system is rated for 16,330 kg load capacity. A ~9,000 kg [10-ton] rated strongback lifting fixture is used to insert the cryomodule onto the isolation fixture. The transport system will be loaded at Fermilab and Jefferson Lab as a unit onto the tractor-trailer. However, the strongback fixture will be used to unload at SLAC.

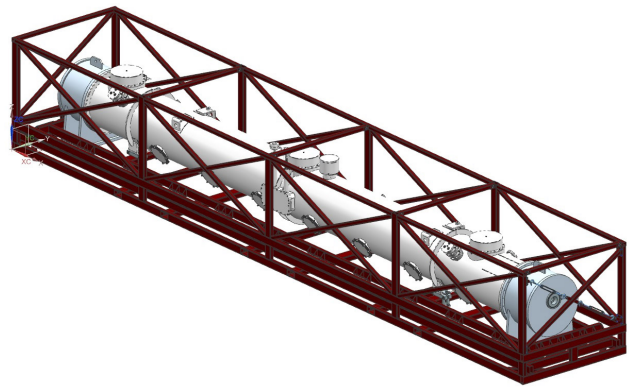

Figure 2: Solid model of cryomodule transport assembly

### Isolation System

The isolation system consisting of a cryomodule, isolation fixture, hardware and transport caps and has a combined weight of 8,910 kg. A system of (32) IDC M28-525-08 isolator (φ22.2 mm cable diameter) helical coils are loaded in compression and rotated at 45 degrees [7]. A design static deflection of 2.54 mm corresponds to an 80% overall the shock reduction, which remains relatively stiff and slightly under-damped.

A serious concern is related to the deflation of the air ride system before or during transport. The suspension system manufacture recommends that Drivers (or Operators) must have limited access to the system, as relieving of the air-ride suspension system hard-wires the transport cargo to the shocks experienced during an over-the-road transport. Automatic exhaust type systems are the worst, as an (inflation or exhaust) cycle can be underway without the Driver's knowledge, and there are no checks and balances to ensure that the air-ride suspension is maintained properly during a random transport.

Our approach involves an independent isolation system built-in to the transport system. Extensive research was completed to understand which type of suspension and trailer to be used. It is proposed to use a 14.6 m [48 ft] air-ride flatbed curtainside trailer for the transports. For example, a common suspension such as the Hendrickson INTRAAX AANT 23K with 9,072 kg per axial load has a natural frequency of 1.26 increasing to 1.5 Hz as the load is increased [8]. The estimated natural frequency of the transport isolation system is 6.3 Hz (and is expected to be uncoupled from the air ride system).

## TRANSPORT CAPS

The cavity string is suspended under the helium gas return pipe (HeGRP), which acts as the beam-line backbone and is supported by three support posts to the vacuum vessel, shown in Figure 3.

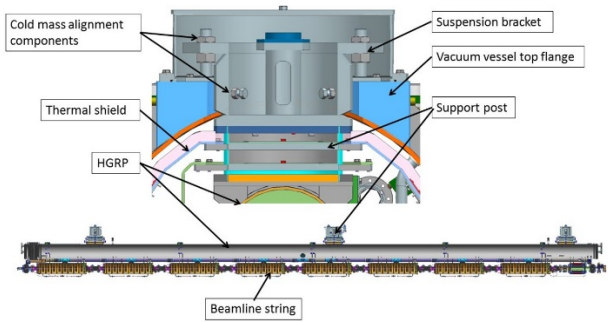

Figure 3: Cold mass support and alignment system [6].

The support and alignment system will provide good cavity, quadrupole, and BPM alignment (<0.5 mm RMS). The alignment and push screws on the suspension brackets provide the positional adjustment of the beamline to the vacuum vessel at room temperature. The adjustment mechanism is enclosed in the insulating vacuum space; the cavity alignment is set relative to the survey fiducials on the vacuum vessel during module assembly and will be maintained during the module thermal and pressure cycling [6].

A feed and end transport cap combination is used to constrain the HeGRP and therefore, the cavity string or coldmass during transport. Constraint of the HeGRP was adopted from the XFEL (DESY) transport design [4]. This design consists of a φ146 mm diameter threaded spindle, driven from each transport cap and a docking feature attachment to the HeGRP (at the opposite end) as shown in Figure 4. Procedurally, a torque of 100 N(m) is applied after securing each transport cap to the vacuum vessel, which equates to an axial load of 4,600 N at the HeGRP, and then locked.

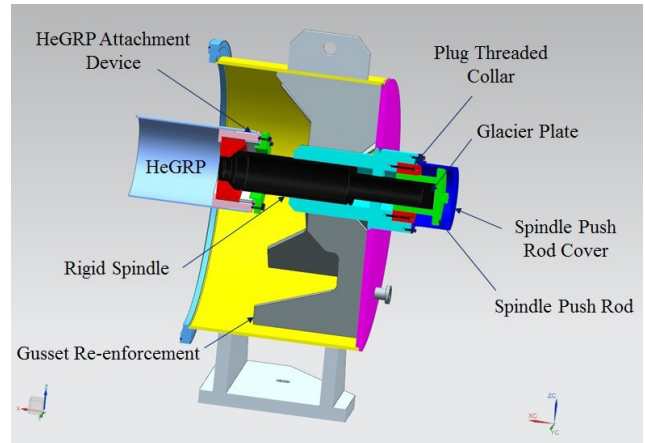

Figure 4: Split solid model of end transport cap.

## TRANSPORT ROUTES AND LOADS

During fair-weather months, a Northern route along I-80 W from Fermilab (3,478 km, 33 hours) and Jefferson Lab (4,788 km, 46 hours) along I-64 W and I-80 W to SLAC will be used. Otherwise, a Southern route along I-40 W from Fermilab (3,858 km, 37 hours) along I-64 W and I-40 W and from Jefferson Lab (4,764 km, 46 hours) will be considered.

### Transport Loads

The effects of shock and vibration are important for both protection and maintaining alignment. Handling operations such as transfers, loading and unloading will be pre-planned and closely monitored, as accidents and miss-handling can cause significant load levels (reaching between 35 to 40 g) [9]. Typical crane and cable systems have a maximum vertical acceleration of 0.6 g. Even when loads are dropped suddenly with a crane, inherent design properties limit the vertical acceleration to 0.94 g [10]. Over-the-road transport speed is limited to 90 km/hr, where loads under normal conditions may reach 4g at the base frame [4].

### Frequency Driven Loads

Response of the tractor-trailer system due to pavement roughness falls into the category of broad-band random vibrations. Generally, tire response to pavement imperfections create resonant modes near 3 Hz. Tractor-trailer systems exhibit bounce, pitch, and roll motions, the trailer is also free to move independently. Longer wheelbases found with tractor-trailer systems, given the lack of a rigid body structure supported by the suspension, allows bending at low frequencies, 5-10 Hz to the first order. Structural bending modes (depending on the trailer) actually occur near 6-7 Hz [11].

To distill the frequency content of the acceleration, the statistical power spectral density (PSD) function is applied. The PSD is the partial derivative of the mean-square value of acceleration with respect to frequency in terms of ($g^2$/Hz) [11]. A PSD threshold for acceleration (shown in Equation 1) of the isolated cryomodule represents the maximum magnitude of acceleration and fatigue failure of internal components [12]. Considering the longest transport duration for Jefferson Lab of 46 hours (at 6.3 Hz), > 1 x $10^6$ cycles may occur if a low frequency resonant condition is present.

$$PSD_{Threshold} = |F_{max}|^2 \left( \frac{(\omega_0^2 - \omega^2)^2 + \left(\frac{\omega_0 \omega}{Q}\right)^2}{k^2 + \left(\frac{m\omega_0 \omega}{Q}\right)^2} \right) \quad (1)$$

## SLAC ACCEPTANCE

In an agreement between Fermilab, Jefferson Lab and SLAC a minimum requirement criteria was established for LCLS-II Cryomodule delivery. Improvements to the Sector 10 access (receiving) area are under consideration in order to facilitate safe and consistent acceptance of cryomodules over the proposed delivery period. This existing SLAC access, 11% grade road shall be paved and straightened, providing proper cryomodule transport access. Also, development of a 21.3 m long x 7.9 m width x 0.36 m thick concrete pad at the entrance and the enlargement of the access door are planned.

The tractor-trailer with cryomodule transport assembly shall be level during unloading and stationary on stable ground. Therefore, a concrete pad is required with a load capacity of ~ 30,000 kg (maximum weight of tractor-trailer with cryomodule transport assembly). The dimensions for a tractor-trailer are 2.6 m width x 3.7 m height x 20.7 m length.

Cavity vacuum will be checked following shipping by means of a vacuum gauge mounted on a beam vacuum manifold at one end of the assembly. Without active pumping and given the particle free environment of the beamline vacuum, the vacuum may rise to a pressure of 1.0 e-4 Torr and stabilize.

### *Alignment*

A post alignment check of (3) Taylor-Hobson fiducials and (2) nests or monuments found on the beam valves is planned. The alignment acceptance criteria is 0.3 mm (vertical and lateral). The longitudinal position will be checked, however there are constraints (column supports) that limit any motion along the beamline axis. The laser tracker system is able to compensate for temperature and humidity effects based on the measurement conditions however, thermal expansion and contraction of the cryomodule components will have an effect. It will be necessary to provide a thermally stable environment within the Sector 10 alignment vault at SLAC in order to minimize measurement error.

## FUTURE WORK

Precautions will be implemented through supervision of the shipment, especially at points of transition. Isolation fixture tuning and transport studies are planned at both Fermilab and Jefferson Lab. Special attention will be placed on understanding any fatigue due to high cycling during transport.

## ACKNOWLEDGEMENTS

Thanks to LCLS-II Cryomodule Fermilab Project Managers; Rich Stanek, Camille Ginsberg and Jay Theilacker. Thanks to Fermilab Mechanical Support Designers; Glenn Waver and Matt Sawtell. We appreciate the expert cryomodule transport advice from Serena Barbanotti and Kay Jensch of DESY. Also, thanks to technical staff consisting of Mark Chlebek, Jim Rife, Glen Smith and Jeff Wittenkeller. Special thanks to Fermilab Staff Al Elste and George Davidson.